\documentclass{article}
\usepackage{spconf,amsmath} %,graphicx}
\usepackage{amssymb,amsfonts}

\usepackage{color}
\usepackage{graphicx}
\usepackage{url}
\usepackage{cite}
\usepackage{setspace}

\newcommand{\vm}[1]{\textit{\boldmath $#1$}}

\title{AttS2S-VC: Sequence-to-Sequence Voice Conversion\\with Attention and Context Preservation Mechanisms}
\name{Kou Tanaka, Hirokazu Kameoka, Takuhiro Kaneko, Nobukatsu Hojo}
\address{NTT Communication Science Laboratories, NTT Corporation, Japan \\
{\small \tt \{tanaka.ko, kameoka.hirokazu, kaneko.takuhiro, hojo.nobukatsu\}@lab.ntt.co.jp} }

\begin{document}
\setlength{\abovedisplayskip}{6pt} % 数式上部のマージン
\setlength{\belowdisplayskip}{6pt} % 数式下部のマージン

%\ninept
%
\maketitle
\begin{abstract}

This paper describes a method based on a sequence-to-sequence learning (Seq2Seq) with attention and context preservation mechanism for voice conversion (VC) tasks.
Seq2Seq has been outstanding at numerous tasks involving sequence modeling such as speech synthesis and recognition, machine translation, and image captioning.
In contrast to current VC techniques, our method 1) stabilizes and accelerates the training procedure by considering guided attention and proposed context preservation losses, 2) allows not only spectral envelopes but also fundamental frequency contours and durations of speech to be converted, 3) requires no context information such as phoneme labels, and 4) requires no time-aligned source and target speech data in advance.
In our experiment, the proposed VC framework can be trained in only one day, using only one GPU of an NVIDIA Tesla K80, while the quality of the synthesized speech is higher than that of speech converted by Gaussian mixture model-based VC and is comparable to that of speech generated by recurrent neural network-based text-to-speech synthesis, which can be regarded as an upper limit on VC performance.

\end{abstract}
\section{Introduction}
\label{sec:intro}

Voice conversion (VC) systems aim to convert para/non- linguistic information included in a given speech waveform while preserving its linguistic information.
VC has been applied to various tasks, such as speaker conversion~\cite{kaneko2017parallel,gao2018voice,kameoka2018stargan} for impersonating or hiding a speaker's identity, as a speaking aid~\cite{kain2007improving,nakamura2012speaking} for overcoming speech impairments, as a style conversion~\cite{inanoglu2009data,toda2012statistical} for controlling speaking styles including emotion, and for pronunciation/accent conversion~\cite{felps2009foreign,kaneko2017sequence} in language learning.

A popular form of VC is a statistical one based on a Gaussian mixture model (GMM)~\cite{toda2007voice}; it requires time-aligned parallel data of the source and target speech for training the conversion models.
For frameworks requiring time-aligned parallel data, other researchers have proposed exemplar-based VCs using non-negative matrix factorization (NMF)~\cite{takashima2013exemplar,wu2014exemplar} and neural network (NN)-based VCs using restricted Boltzmann machines~\cite{chen2014voice,nakashika2014voice}, feed-forward NNs~\cite{desai2010spectral,saito2017voice}, recurrent NNs~\cite{sun2015voice,nakashika2014high}, variational autoencoders~\cite{hsu2016voice,saito2018non}, and generative adversarial nets~\cite{kaneko2017sequence}.
On the other hand, frameworks requiring no parallel data, called parallel-data-free VCs, have been proposed~\cite{kaneko2017parallel,kameoka2018stargan} to avoid the time-consuming job of recording speech for parallel data collection.
Notably, the drawbacks of these VCs are the prerequisite of a large number of transcripts and/or difficulty converting the durations of the source speech.

Recently, sequence-to-sequence (Seq2Seq) learning~\cite{sutskever2014sequence,bahdanau2014neural} has proved to be outstanding at various research tasks such as text-to-speech synthesis (TTS) ~\cite{wang2017tacotron,shen2018natural,tachibana2018efficiently} and automatic speech recognition (ASR)~\cite{chan2016listen,chiu2018state}.
The early Seq2Seq model~\cite{sutskever2014sequence} has encoder and decoder architectures for mapping an input sequence to an encoded representation used by the decoder network to generate an output sequence.
To select critical information from the encoded representation in accordance with the output sequence representation, later Seq2Seq models~\cite{bahdanau2014neural,raffel2017online} introduce an attention mechanism.
The key advantages of the Seq2Seq learning approach are the ability to train a single end-to-end model directly on the source and target sequences and the capacity to handle input and output sequences of different lengths.
In particular, we expect that the Seq2Seq model makes it possible to convert not only acoustic features but also the durations of the source speech to those of the target speech.
Moreover, Seq2Seq learning is extensible to semi-supervised learning~\cite{andros2017listening}, where it can avoid the time-consuming task of collecting parallel data.
In a supervised learning task, Seq2Seq learning requires parallel data of the source and target sequences rather than time-aligned parallel data.
Considering dual learning~\cite{he2016dual,zhu2017unpaired}, Seq2Seq learning can be trained with a small amount of parallel data and a large amount of non-parallel data.

In this paper, we propose a Seq2Seq-based VC with attention and context preservation mechanisms\footnote{\label{foot1}Audio samples can be accessed on our web page: \url{http://www.kecl.ntt.co.jp/people/tanaka.ko/projects/atts2svc/attention_based_seq2seq_voice_conversion.html}}.
Our contributions are as follows:
\begin{itemize}
\setlength{\itemsep}{-0.0mm}
\setlength{\parskip}{-0.0mm}
\item Our VC method makes it possible to stabilize and accelerate the training procedure by considering guided attention and context preservation losses.
\item It makes it possible to convert not only spectral envelopes but also fundamental frequency contours and durations of the speech.
\item It requires no context information such as phoneme labels, unlike~\cite{miyoshi2017voice,zhang2018sequence} which introduced the Seq2Seq model and used context information.
\item It requires no time-aligned source or target speech data in advance.
\end{itemize}
We conducted an our experiment demonstrating that the quality of the synthesized speech generated by our VC framework is higher than that of speech generated by the conventional GMM-based VC, and it is comparable to that of speech generated by recurrent-NN based TTS in terms of both naturalness and speaker similarity.
Note that the proposed model was trained in only one day, using only one GPU of an NVIDIA Tesla K80.

\section{Conventional VC}

\subsection{Frame/Sequence- based VC}
\label{sec:VCbaseline1}

There are two types of frame/sequence- based VC: VCs requiring parallel data~\cite{stylianou1998continuous,toda2007voice,helander2010voice} and parallel-data-free VC~\cite{kaneko2017parallel,kameoka2018stargan}.
The first framework has different procedures of training and conversion, as shown in Fig.~\ref{fig:conventionalVC}a.
The conversion procedure does not have a time warping function, despite that the training procedure includes a time-alignment step to handle source and target sequences having different lengths.
The second framework is a parallel-data-free VC that does not require parallel source and target speech data.
To realize parallel-data-free VC, the second framework uses context information~\cite{lifa2016phonetic,liu2018wavenet}, adaptation techniques~\cite{mouchtaris2006nonparallel,lee2006map}, a pre-constructed speaker space~\cite{toda2006eigenvoice,saito2011one}, and cycle consistency~\cite{kaneko2017parallel,kameoka2018stargan}.
Although these VC techniques have various training procedures, the conversion procedure does not involve the time warping function.
Consequently, the frame/sequence- based VC frameworks do not allow us to convert the durations and the acoustic features of the source speech at the same time.
In contrast, our model allows both the acoustic features and the durations to be converted at the same time.

\subsection{Seq2Seq-based VC}
\label{sec:VCbaseline2}

\begin{figure}[!t]
    \centering
    \includegraphics[width=80mm,clip]{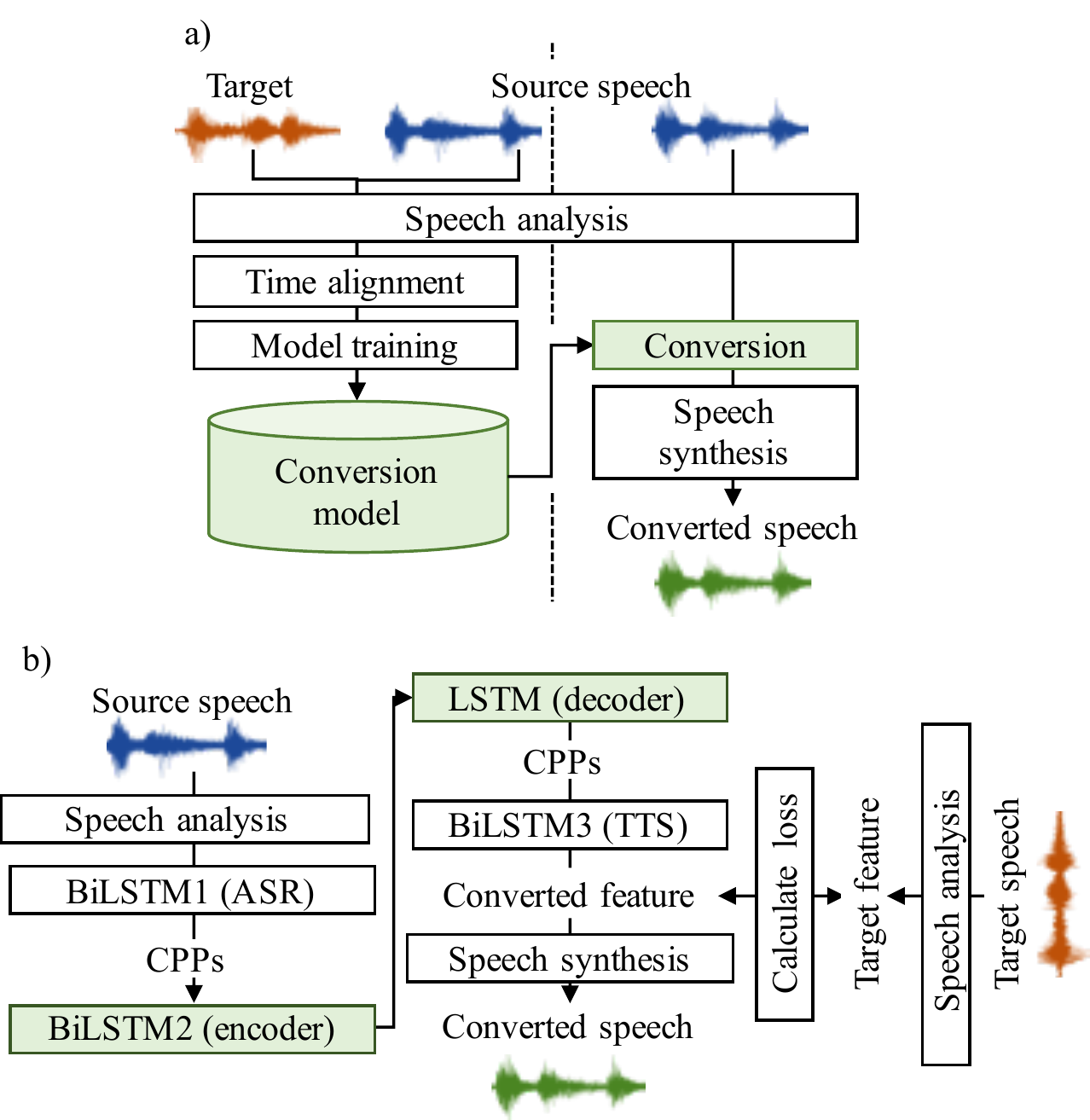}
    %\includegraphics[width=80mm,clip]{conventionalVC.pdf}
    %\vspace{-3mm}
    \caption{System overviews of conventional VC, a) frame/sequence- based VC using parallel data (see Sec.~\ref{sec:VCbaseline1}) and b) Seq2Seq-based VC (see Sec.~\ref{sec:VCbaseline2}). ``CPPs'' and ``BiLSTM'' denote context posterior probabilities and bidirectional LSTM, respectively.}
    \label{fig:conventionalVC}
    %\vspace{-3mm}
\end{figure}

In contrast to the frame/sequence- based VC frameworks, Seq2Seq-based VC frameworks make it possible to convert not only the acoustic features but also the durations of the source speech.
Most Seq2Seq-based VCs consist of ASR and TTS modules which are trainable with pairs of speech and its transcript rather than the source and target speech.
The ASR module converts the acoustic feature sequence of the source speech into a sequence of context information such as phoneme labels and context posterior probabilities~\cite{miyoshi2017voice,liu2018wavenet}, and the TTS module generates the acoustic feature sequence of the desired speech from the sequence of context information.
One approach to changing the duration of the source speech uses a re-generation method that generates the duration information from the text symbols after converting the acoustic features of the source speech into text symbols once.
Namely, the duration information is erased once and re-generated.
Another approach~\cite{miyoshi2017voice,zhang2018sequence} involves Seq2Seq learning, as shown in Fig.~\ref{fig:conventionalVC}b.
In this approach, the context posterior probability sequence of the source speech including the duration information is directly converted into a context posterior probability sequence of the desired speech including the duration information.
Both approaches work well if ASR performs robustly and accurately enough, but they require a large number of transcripts to train each module.
In contrast, our model does not use any transcript.

\section{AttS2S-VC}

\begin{figure}[!t]
    \centering
    \includegraphics[width=80mm,clip]{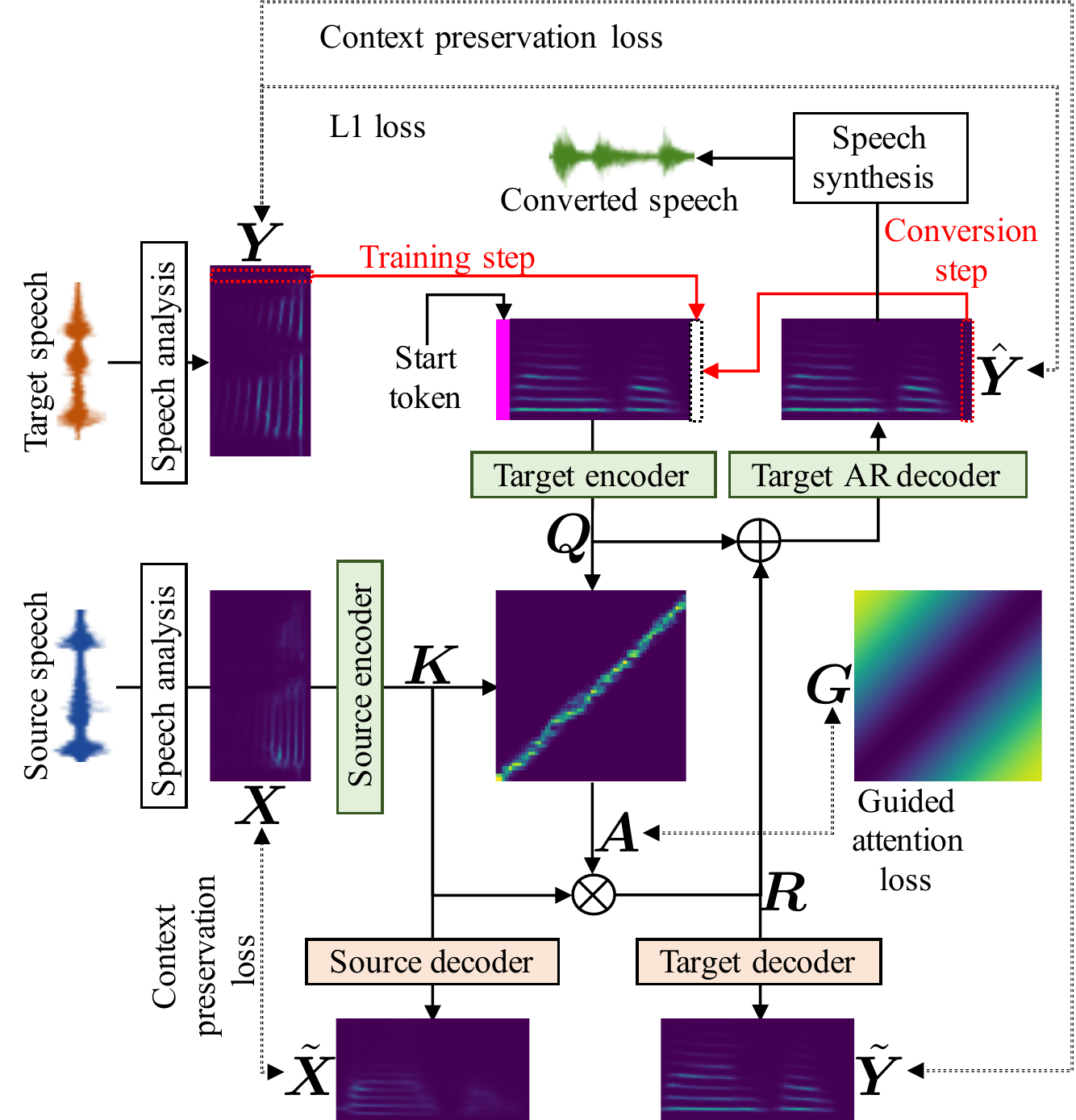}
    %\vspace{-3mm}
    %\includegraphics[width=80mm,clip]{conventionalVC.pdf}
    \caption{System overviews of proposed VC. The solid lines indicate the training and conversion procedures. The dashed lines indicate calculations of the differences during training. Green boxes and red boxes respectively denote the original components and the proposed additional components.}
    \label{fig:proposedVC}
    %\vspace{-3mm}
\end{figure}

Our method consists of 1) four basic components of the Seq2Seq model and 2) two additional components as a context preservation mechanism.
The four basic components are a source encoder, target encoder, target autoregressive (AR) decoder, and attention mechanism.
The two additional components are a source decoder and another target decoder to keep linguistic information of the source speech.
Figure~\ref{fig:proposedVC} is an overview of the system.

\subsection{Seq2Seq Model with Attention Mechanism}

Let us use $\vm{X} = \left[\vm{x}_1, \cdots, \vm{x}_I \right]$ and $\vm{Y} = \left[\vm{y}_1, \cdots, \vm{y}_J \right]$ to denote sequences of acoustic features of the source and target speech, respectively.
The source encoder network $f_\mathrm{SrcEnc}$ and target encoder network $f_\mathrm{TarEnc}$ encode the input sequences $\vm{X}$ and $\vm{Y}$ to the embeddings $\vm{K} = \left[\vm{k}_1, \cdots, \vm{k}_I \right]$ and $\vm{Q} = \left[\vm{q}_1, \cdots, \vm{q}_J \right]$, as follows:
\begin{align}
    \vm{K} = f_\mathrm{SrcEnc}(\vm{X}), \\
    \vm{Q} = f_\mathrm{TarEnc}(\vm{Y}).
\end{align}

In order to accurately predict the output sequence, ~\cite{bahdanau2014neural,raffel2017online} introduced an attention mechanism.
At each time frame of the embeddings $\vm{Q}$, the attention mechanism gives a probability distribution that describes the relationship between the given time frame feature $\vm{q}_j$ and the embeddings \vm{K}.
Consequently, the attention matrix $\vm{A}$ can be written as
\begin{align}
    e_{i,j} = f_\mathrm{FFNN}(\vm{k}_i, \vm{q}_j), \\
    a_{i,j} = \frac{\exp(e_{i,j})}{\sum_i \exp(e_{i,j})}.
\end{align}
\noindent where $f_\mathrm{FFNN}$ indicates a function described by feed-forward NNs and $a_{i,j}$ is an element ($i$, $j$) of the attention matrix $\vm{A}$.

A seed $\vm{R} = \left[\vm{r}_1, \cdots, \vm{r}_J \right]$ of the target AR decoder is obtained by considering the long-range temporal dependencies between the source and target sequences as follows:
\begin{align}
    \vm{r}_j = \vm{K}\vm{a}_j.
\end{align}
\noindent As the name implies, the target AR decoder involves all previous outputs of itself.
Hence, the input of the target AR decoder is $\vm{R}'$ combined with the seed $\vm{R}$ and the embeddings $\vm{Q}$.
The output $\hat{\vm{Y}}$ of the Seq2Seq model is obtained through the target AR decoder $f_\mathrm{TarDecAR}$,
 \begin{align}
    \hat{\vm{Y}} = f_\mathrm{TarDecAR}(\vm{R}').
\end{align}
\noindent Finally, we minimize the objective function ${\mathcal L}_\mathrm{Seq2Seq}$ of Seq2Seq learning: 
\begin{align}
    {\mathcal L}_\mathrm{Seq2Seq} = || \hat{\vm{Y}} - \vm{Y} ||_1.
\end{align}

\subsection{Stabilizing and Accelerating Training Procedure}

\subsubsection{Guided Attention Loss}

To accelerate the training of an attention module, ~\cite{tachibana2018efficiently} introduced a guided attention loss.
Generally speaking, most speech signal processing applications, such as ASR, TTS, and VC, are time incremental algorithms.
It is natural to assume that the time frame $i$ of the source speech waveform progresses nearly linearly with respect to the time frame $j$ of the target speech waveform, i.e., $i \sim \alpha j$, where $\alpha \sim \frac{I}{J}$.
Therefore, the attention matrix $\vm{A}$ should be a nearly diagonal.
A penalty matrix $\vm{G}$ is designed, as follows:
\begin{align}
    g_{i,j} = 1 - \exp \left\{ \frac{ - (\frac{i}{I} - \frac{j}{J})^2}{2 \sigma_g^2} \right\},
\end{align}
\noindent where $\sigma_g$ controls how close $\vm{A}$ is to a diagonal matrix.
The guided attention loss ${\mathcal L}_\mathrm{ga}$ is defined as
\begin{align}
    {\mathcal L}_\mathrm{ga} = || \vm{G} \odot \vm{A} ||_1,
\end{align}
\noindent where $\odot$ indicates an element-wise product.

\subsubsection{Context Preservation Loss}

To stabilize the training procedure, we propose a context preservation loss.
In preliminary experiments, we found that the training procedure sometimes failed even if it took into account the guided attention loss (see speech samples on our web page~\ref{foot1}).
In particular, the converted speech seemed like randomly generated speech or speech repeating several phonemes.
One possible reason is that minimizing the objective function ${\mathcal L}_\mathrm{Seq2Seq}$ sometimes makes the target AR decoder a network just reconstructing the input of the target encoder.
It is because we use $\vm{Y}$ rather than $\hat{\vm{Y}}$ as the input of the target encoder in the training.
As a result, the source encoder is not required to control the output of the target AR decoder and preserve the context information of the source speech.

To make the source encoder meaningful, we introduce two additional networks to the original Seq2Seq model as a context preservation mechanism.
One is a source decoder $f_\mathrm{SrcDec}$ for reconstructing the source speech $\tilde{\vm{X}}$ from the embeddings $\vm{K}$.
The other is a target decoder $f_\mathrm{TarDec}$ for predicting the target speech $\tilde{\vm{Y}}$ from the seed $\vm{R}$.
 \begin{align}
    \tilde{\vm{X}} = f_\mathrm{SrcDec}(\vm{K}), \\
    \tilde{\vm{Y}} = f_\mathrm{TarDec}(\vm{R}).
\end{align}
\noindent From another point of view, the source decoder $f_\mathrm{SrcDec}$ helps the source encoder to preserve the linguistic information of the source speech $\vm{X}$, while the target decoder $f_\mathrm{TarDec}$ helps the source encoder to encode the source speech $\vm{X}$ to the shared space of the source and target speech.
Note that in the preliminary experiments, the target decoder was more important than the source decoder.
The full objective function of our model is formulated as
\begin{align}
    {\mathcal L}_\mathrm{proposed} & = {\mathcal L}_\mathrm{Seq2Seq} + \lambda_\mathrm{ga} {\mathcal L}_\mathrm{ga} \nonumber \\ 
    & + \lambda_\mathrm{cp} (|| \tilde{\vm{X}} - \vm{X} ||_1 + || \tilde{\vm{Y}} - \vm{Y} ||_1),
\end{align}
\noindent where $\lambda_\mathrm{cp}$ controls the context preservation loss.

\section{Experiments}

\subsection{Experimental Conditions}

\noindent {\bf Datasets:}
We used the CMU Arctic database~\cite{kominek2004cmu} consisting of utterances by two male speakers ({\bf rms} and {\bf bdl}) and two female speakers ({\bf clb} and {\bf slt}).
To train the models, we used about 1,000 sentences (speech section of 50 min) of each speaker.
To evaluate the performance, we used 132 sentences of each speaker.
The sampling rate of the speech signals was 16 kHz.
We treated {\bf rms} and {\bf clb} as source speakers and {\bf bdl} and {\bf slt} as target speakers.
For the evaluations, we conducted intra-gender pairs, {\bf rms}-{\bf bdl} and {\bf clb}-{\bf slt}, and cross-gender pairs, {\bf rms}-{\bf slt} and {\bf clb}-{\bf bdl}.
Note that we trained the conversion models for every speaker pair, independently.

{\bf Baseline system 1 (GMM-VC-wGV):}
We used a GMM-based VC method~\cite{toda2007voice} as the baseline for frame/sequence- based VC described in Sec.~\ref{sec:VCbaseline1}.
To train the conversion models, we used an open source VC toolkit sprocket~\cite{kobayashi2018sprocket} and its default settings, except for $F_0$ ranges and power thresholds.
Note that a global variance (GV)~\cite{toda2007voice} was also considered.

{\bf Baseline system 2 (LSTM-TTS):}
By assuming the ASR module and the encoder part of the encoder-decoder module in~\cite{miyoshi2017voice} work perfectly, we can focus on the TTS module.
Therefore, we used an LSTM-based TTS method as the baseline of Seq2Seq-based VC described in Sec.~\ref{sec:VCbaseline2}.
The contextual features used as input were 416-dimensional linguistic features obtained using the default question set of the open source TTS toolkit Merlin~\cite{wu2016merlin}.
From the speech data, 60 Mel-cepstral coefficients, logarithmic $F_0$, and coded aperiodicities were extracted every 5 ms with the WORLD analysis system~\cite{morise2016world}.
As the duration model, we stacked three LSTMs with 256 cells followed by a linear projection.
As the acoustic model, we stacked three bidirectional LSTMs with 256 cells followed by a linear projection.

{\bf Proposed system (Proposed):}
Inspired by Tacotron~\cite{wang2017tacotron}, we used the architecture described in open Tacotron~\cite{keithURLtacotoron}.
Note that we replaced all ReLU activations~\cite{nair2010rectified} with a gating mechanism of gated linear units~\cite{dauphin2016language}.
Although the proposed method worked well for not only acoustic features of the WORLD vocoder but also raw spectral features, we chose to use the acoustic features of WORLD vocoder to balance the experimental conditions of {\bf LSTM-TTS}.
Note that the target AR decoder also generated the stop tokens.
As the additional source decoder and target decoder networks, we used the same architectures as in the source encoder.
The hyperparameters $\sigma_g$, $\lambda_\mathrm{ga}$, $\lambda_\mathrm{cp}$ were 0.4, 10,000, and 10, respectively.  
The batch size, number of epochs, and reduction factor~\cite{zen2016fast} were 32, 1,000 and 5.
We used the Adam optimizer~\cite{kingma2014adam} and varied the learning rate over the course of training~\cite{ashish2017ashish}.

\subsection{Experimental Results}

\begin{figure}[!t]
    \centering
    \includegraphics[width=\columnwidth,clip,trim=0mm 32mm 0mm 0mm]{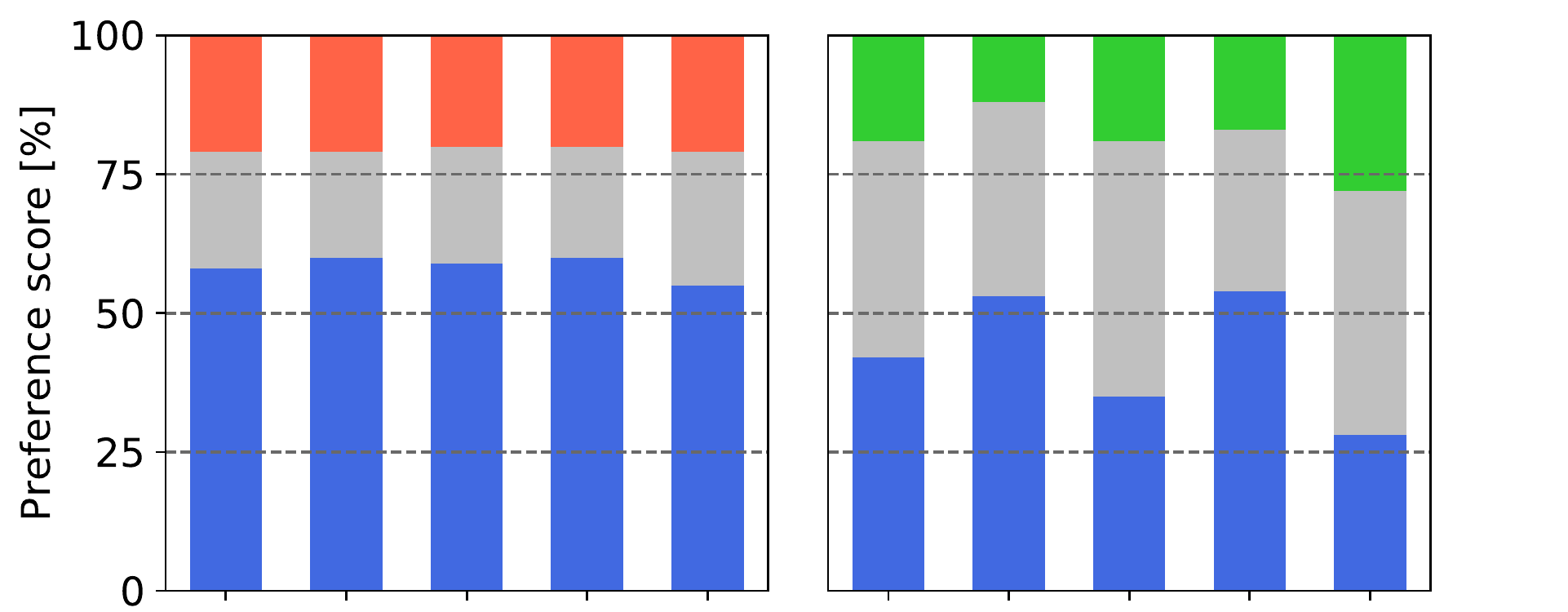}
    \centering
    \includegraphics[width=\columnwidth,clip]{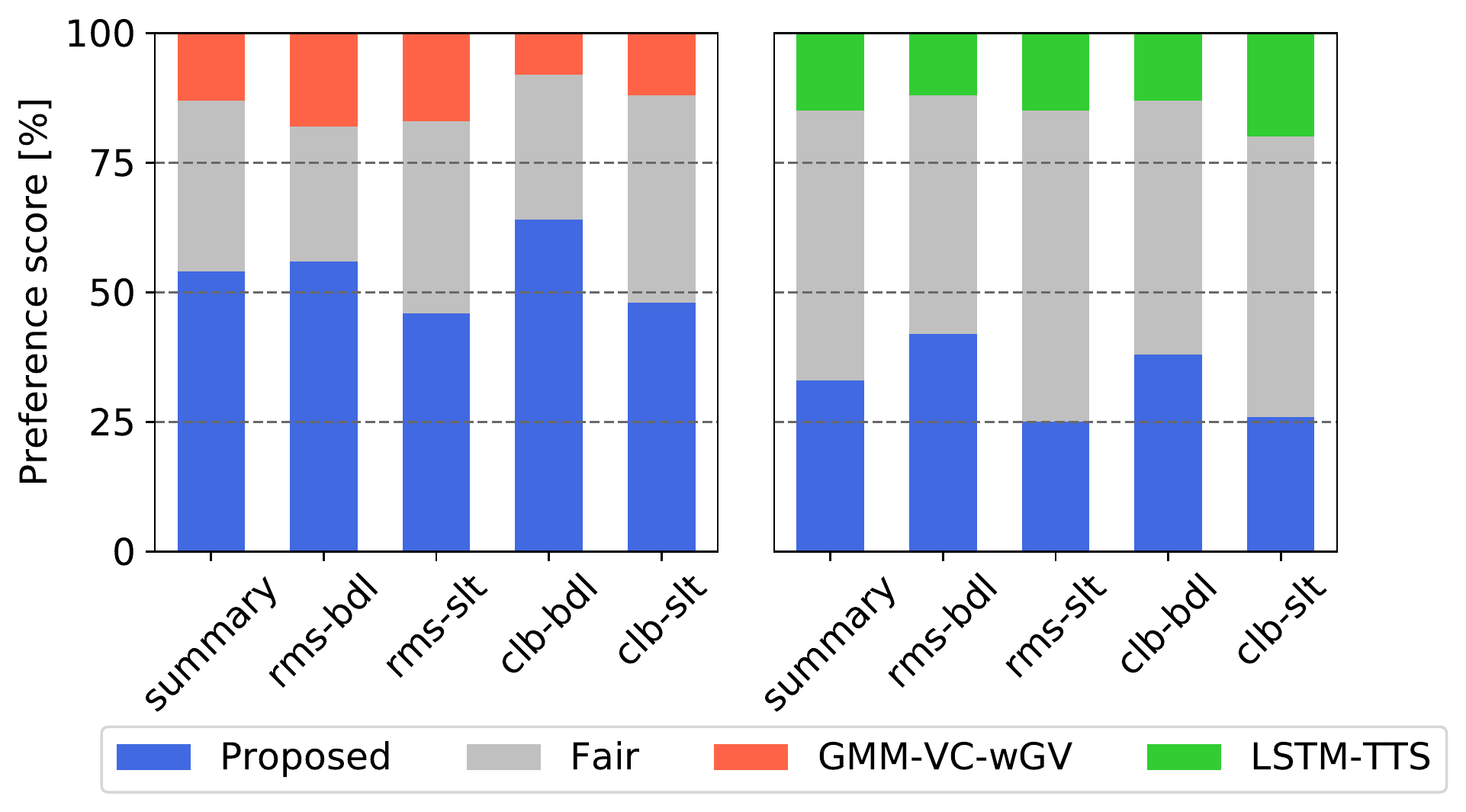}
    \caption{Results of preference tests of naturalness (upper) and speaker similarity (lower).}
    \label{fig:preferencetest}
    %\vspace{-3mm}
\end{figure}

As shown in Fig.~\ref{fig:preferencetest}, we conducted two subjective evaluations, preference tests on naturalness and speaker similarity.
The number of listeners was 15, and each listener evaluated 80 shots consisting of randomly selected 10 speech samples $\times$ 4 pairs of intra/cross- gender $\times$ 2 comparisons, v.s. {\bf GMM-VC-wGV} and v.s. {\bf LSTM-TTS}.

The evaluations indicated that {\bf Proposed} outperformed {\bf GMM-VC-wGV} in terms of both naturalness and speaker similarity.
This is because our method makes it possible to convert not only the acoustic features but also the durations of speech.
In contrast, baseline system 1 forces the conversion while preserving the durations of the source speech.
Consequently, durations not used in the target speech make the conversion errors larger.

Moreover, {\bf Proposed} was comparable to {\bf LSTM-TTS}.
This result demonstrates that our method makes it possible to learn the key components for changing the individuality of the speaker while preserving the linguistic information.
Notably, our model was trained without any transcript while ~\cite{miyoshi2017voice,zhang2018sequence} used a large number of transcripts.

\section{Conclusions}

We proposed a method based on Seq2Seq learning with attention and context preservation mechanisms for VC tasks.
Experimental results demonstrated that the proposed method outperformed the conventional GMM-based VC and was comparable to LSTM-based TTS.
Extending the proposed method so that it can be used in semi-supervised learning tasks is ongoing work.
Note that since we also progressed in a convolutional version of the proposed method~\cite{kameoka2018convs2svc} simultaneously, we will conduct further evaluations and report the results.

\noindent {\bf Acknowledgements:} This work was supported by a grant from the Japan Society for the Promotion of Science (JSPS KAKENHI 17H01763).

\vfill\pagebreak

%\section{REFERENCES}
%\label{sec:refs}


%\end{spacing}

\end{document}